\documentstyle[aps,prl,floats,twocolumn,psfig]{revtex}
\begin{document}
\draft
\title{Interaction-induced delocalization of quasiparticle pairs in
  the Anderson insulator} 
\author{Felix von Oppen\footnote{Present address: Dept.\ of Condensed
Matter Physics, Weizmann Institute of Science, 76100 Rehovot, Israel}
 and Tilo Wettig}
\address{Max-Planck-Institut f\"ur Kernphysik, 69117 Heidelberg,
  Germany} 
\date{November 7, 1995} 

\twocolumn[
\maketitle\widetext\vspace*{-5mm}\leftskip=1.9cm\rightskip=1.9cm
  It has recently been established that  a short-range interaction can
  strongly delocalize  a   pair of particles   moving  in a disordered
  potential.   We investigate whether an  analogous effect exists also
  for  pairs of quasiparticles in  the Anderson insulator by employing
  an   approximate   numerical evaluation of    the two-particle Green
  function  of  the  many-body system.    At   the Fermi energy,   the
  quasiparticle pair is localized  on the scale of the single-particle
  localization length.  The  delocalization effect  is recovered  when
  the excitation energy of the pair is comparable to the bandwidth.
\begin{abstract}
\\\vspace*{-9mm}
\end{abstract}]

\narrowtext 

The localization  of particles by  a random potential has been studied
extensively over the  past decades \cite{Lee}. While wavefunctions are
exponentially localized in one and two dimensions even for arbitrarily
weak  disorder, there    exists  a transition   between  extended  and
localized  states  in three dimensions.   A non-interacting degenerate
Fermi  gas in a  disordered potential   is conducting  when the  Fermi
energy lies in the region of extended states (metallic phase) while it
is insulating when  the Fermi energy  lies in the region  of localized
states (Anderson insulator).   The situation becomes much more complex
and    controversial   once interactions  between   the  particles are
included.  Interactions may lead to a metal-insulator transition (Mott
transition) even in the absence of  disorder; they also result in rich
magnetic behavior    \cite{Belitz}.   Interest in   the  interplay  of
disorder and interactions was recently  renewed in part by experiments
measuring persistent currents  in mesoscopic normal-metal rings  whose
amplitude  could  not  be  explained  by theories for  non-interacting
electrons \cite{Riedel}.

An  original approach to   studying  localization in the  presence  of
interactions was recently   taken by Shepelyansky  \cite{Shepelyansky}
and previously by Dorokhov  \cite{Dorokhov}.  They considered the much
simpler problem of two interacting particles in a random potential and
predicted  that      the  interaction can  lead      to  a significant
delocalization of the pair.  Shepelyansky \cite{Shepelyansky} and Imry
\cite{Imry} speculated that an  analogous effect could exist for pairs
of   quasiparticles  at  finite    particle  density  which  may  have
significant   consequences  for   the   thermodynamic and    transport
properties  of  the Anderson insulator.     It is the   purpose of the
present paper  to study the propagation  of quasiparticle pairs in the
Anderson insulator in the presence of a short-range interaction within
a   numerically tractable  approximation    to the two-particle  Green
function.   Our approximation  is   motivated   by both  physical  and
analytical considerations.
 
We briefly recall  some of the pertinent  results for the two-particle
problem. Shepelyansky \cite{Shepelyansky}  studied two particles in  a
random  potential interacting by  a short-range interaction.  Whenever
the two    particles   are  localized  far  apart   compared   to  the
single-particle  localization  length    $\xi_1$, the   effect of  the
interaction is   only exponentially   small.    By contrast,   the two
particles can propagate as a pair over a distance $\xi_2$ which can be
much larger  than $\xi_1$  when  they are localized  within about  one
single-particle  localization  length of   each  other.  Specifically,
Shepelyansky    predicted     by  an   approximate   mapping     to  a
banded-random-matrix  model  which  he  studied  numerically that  the
two-particle      localization     length     $\xi_2$        satisfies
$\xi_2/\xi_1\simeq(\xi_1/32)(u/t)^2$, where   $u$ denotes    the   disorder
strength and $t$ the bandwidth (hopping matrix element). The effect is
essentially  independent of  the   sign  of the interaction    and the
statistics of the    particles.  Subsequently, the existence  of  this
effect  was  confirmed  using both a    Thouless-type scaling argument
\cite{Imry} and  numerical calculations \cite{Frahm,Oppen}.  Numerical
work \cite{Oppen} computing  $\xi_2$ directly from  the Green function
of the two-particle problem showed  that in strictly one dimension the
two-particle localization length at the center of the band satisfies a
scaling relation  $\xi_2/\xi_1=f(u\xi_1/t)$  with $f(x)\simeq1/2+C|x|$
which, while qualitatively confirming the effect, is inconsistent with
the original prediction mentioned above ($C$ is a numerical constant).

The most important additional  feature of the degenerate  Fermi system
is the Pauli  principle.   Two particles  propagating in a  disordered
potential sufficiently far from the band edges  have ample phase space
for scattering.   On the  other hand, phase  space for   scattering is
severely restricted at finite density since all states below the Fermi
energy are effectively  blocked.  One may expect that  this leads to a
suppression of the delocalization at low pair excitation energy.  With
increasing excitation energy, the phase-space restriction becomes less
severe  and the delocalization  effect should be  recovered.  We study
the   influence of  the    Pauli  principle  on  the  propagation   of
quasiparticle  pairs within  an approximation analogous  to the Cooper
problem of  superconductivity: We consider    two particles above  the
Fermi   energy which  interact  with  each   other,  but not with  the
electrons in the  Fermi sea except via the   exclusion principle.  The
Fermi sea is defined by filling up  single-particle eigenstates of the
disorder potential below the Fermi  energy $E_F$.   In order to  avoid
the   superconducting   instability,   we  consider   only   repulsive
interactions.

One can motivate this approximation also by  a quite different line of
thought.  It is plausible that the two-particle  effect is most likely
to survive at low   particle density.  When considering a  short-range
interaction  in three  dimensions, only  ladder  diagrams involving no
intermediate  hole  excitations  contribute  to leading   order in the
low-density limit \cite{Fetter}.  Hence, the Cooper problem is in fact
equivalent to  a  low-density approximation.  This   can be understood
physically  by considering  some second-order   diagrams as shown   in
Fig.~1.  While the ladder diagram in  Fig.~1(a) involves only particle
excitations   at the intermediate stage,  the   diagrams 1(b) and 1(c)
involve   intermediate hole   excitations.   The  latter  diagrams are
suppressed  at low   density because  they involve  particle-hole-pair
creation.  For these  reasons,  we will only  study  particle-particle
excitations  in   this paper.   We    expect that  our   results apply
qualitatively also to hole-hole and particle-hole excitations.
\begin{figure}
\centerline{\psfig{figure=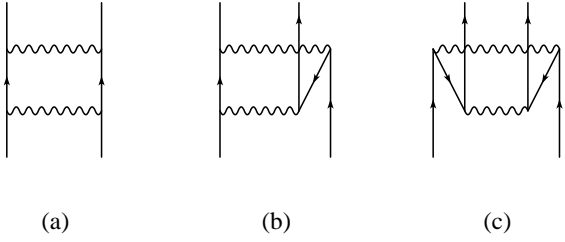,width=75mm}}
\caption{Examples of elementary second-order diagrams in the 
  particle-particle channel with  time running upwards.  While diagram
  (a) contributes to leading  order in the low-density limit, diagrams
  (b)  and (c)  are  suppressed since  they involve particle-hole-pair
  creation and annihilation.}
\end{figure}

We  now   proceed to  work out   the Cooper   approximation within the
diagrammatic   approach.    We   consider  spinless  fermions     on a
one-dimensional lattice with repulsive nearest-neighbor interaction of
strength $u$ and subject to a random potential,
\begin{eqnarray}
  {\hat H}&=&t\sum_n [{\hat a}_n^\dagger {\hat a}_{n+1}+
  {\hat a}_{n+1}^\dagger{\hat a}_n]+\sum_n v_n
  {\hat a}_n^\dagger {\hat a}_n \nonumber \\
  &&+u\sum_n {\hat a}_n^\dagger {\hat a}_{n+1}^\dagger 
  {\hat a}_{n+1}{\hat a}_n\;.
\label{Hamiltonian} 
\end{eqnarray}
Here, ${\hat a}_n$ denotes the  fermionic  annihilation operator of  a
particle  at site   $n$,  and $v_n\in[-W/2,W/2]$  denotes  the  random
on-site energies. In the following, energies will be measured in units
of the hopping matrix element $t$ and lengths in  units of the lattice
spacing   $a$.   As  motivated   above,    we focus on    the retarded
two-particle  Green    function  in the    particle-particle  channel.
Analogous to our approach \cite{Oppen} to the two-particle problem, we
consider  only processes where the    quasiparticles  are created   on
neighboring sites  $(m,m+1)$ at time  $\tau=0$  and destroyed  at time
$\tau$   on a   different set of   neighboring  sites  $(n,n+1)$.  The
corresponding amplitude is given by
\begin{eqnarray}
  F^{\rm pp}(n,m;\tau)&=&-i\theta(\tau)\langle0|{\hat a}_n(\tau){\hat
    a}_{n+1}(\tau) {\hat a}_{m+1}^\dagger(0){\hat
    a}_m^\dagger(0)|0\rangle\nonumber\\ 
  &=&\int_{-\infty}^\infty{dE\over2\pi}\,F^{\rm pp}(n,m;E)
  \exp(-iE\tau/\hbar).
\end{eqnarray}
The localization length $\xi_q$  for    coherent propagation of    the
quasiparticle pair can be defined by the exponential decrease of these
matrix elements with distance,
\begin{equation}
  {1\over\xi_q(E)}=-\lim_{|n-m|\to\infty}{1\over|n-m|}\langle
        \ln|F^{\rm pp}(n,m;E)|
    \rangle\;,
\end{equation}
where $\langle\ldots\rangle$  denotes   a  disorder   average.    When
considering the nearest-neighbor interaction in (\ref{Hamiltonian}), a
closed equation  can    be derived for  the   matrix  elements $F^{\rm
  pp}(n,m;E)$  within the  ladder approximation.  It  is this enormous
simplification which makes our numerical calculations feasible.

Computing    $F^{\rm pp}(n,m;E)$   for non-interacting  fermions,  one
readily obtains
\begin{equation}
  F_0^{\rm pp}(n,m;E)=\sum_{\epsilon_i>\epsilon_j>E_F}{B^*_{i,j}(n)
    B_{i,j}(m)\over E-\epsilon_i-\epsilon_j+i\eta}\;.
\label{green0}
\end{equation}
Here, $\epsilon_i$   denotes    the  single-particle energy  of    the
eigenstate $\phi_i(n)$ of the disorder potential, $\eta$ is a positive
infinitesimal, and $B_{i,j}(n)=    \phi_i(n)\phi_j(n+1)-   \phi_i(n+1)
\phi_j(n)$.   Summing  the ladder diagrams   and considering  only the
contributions of  intermediate particle-particle states,  one  obtains
the two-particle Green function in the Cooper approximation,
\begin{equation}
  F^{\rm pp}(E)={F_0^{\rm pp}(E)\over u}{1\over1/u-F_0^{\rm pp}(E)}\;,
\label{green}
\end{equation}
where $F^{\rm  pp}(E)$ is  viewed as  a matrix in  site space.   It is
instructive  to compare these equations  to those for the two-particle
problem \cite{Oppen}.  The  difference between the  two cases is  that
the sum in  (\ref{green0}) is only over  states above the Fermi energy
while the corresponding equation for the two-particle case involved an
unrestricted summation over all states.   It is possible to extend our
approach to   the full ladder approximation.  Then, Eq.~(\ref{green0})
would also contain a  summation over hole-hole excitations.   However,
we decided to use the Cooper approximation because it can be motivated
systematically in  the low-density limit.   We do not  expect that our
results change qualitatively when hole-hole excitations are included.

The first factor on the right-hand side  of (\ref{green}) decreases on
the scale of the single-particle localization length.  Hence, we focus
on the second factor in    the following  from which any    long-range
behavior  must arise.  For  numerical  purposes, it  is instructive to
interpret this  term  as the Green   function  for the ``Hamiltonian''
$F_0^{\rm pp}(E)$ at ``energy''  $1/u$.  Since $F_0^{\rm pp}(E)$  is a
banded matrix whose bandwidth  is of the  order of the single-particle
localization length, this allows us to compute the localization length
$\xi_q$ by employing  the efficient recursive Green-function technique
for banded Hamiltonian matrices \cite{Huckestein}.
 
For   our numerical  calculations  we  chose a  Fermi energy $E_F=-1$,
corresponding  roughly to  quarter    filling.  We have  studied   the
localization length of quasiparticle pairs for nine values of the pair
energy $E$ ranging  from $E=-2$ to $E=0$  in steps of $\Delta E=0.25$.
This   corresponds to excitation   energies  $\epsilon=E-2E_F$  of the
quasiparticle pair between  $\epsilon=0$ and  $\epsilon=2$.  For  each
value  of   the excitation energy we    used three values  of disorder
$W=1.5,2$,  and $3$,   corresponding  to  single-particle localization
lengths at   the   band center    $\xi_1=46.7,   26.2$,  and   $11.7$,
respectively, and five  values of the interaction $u=0.1,0.3,0.5,0.75$
and  $1.0$.   For comparison,  we have  also   computed $\xi_2$ of the
two-fermion problem for the  same set of parameters.  The computations
were done   for  lattices with 500    sites, and we  averaged over  50
realizations of  the  disorder.  We briefly  comment  on our choice of
$E_F$.   As argued above,  our approximation can  be  motivated by the
low-density  limit.  In view  of  this, our choice   of $E_F$ may seem
quite large.   We chose this somewhat larger  value of $E_F$ since the
two-particle effect becomes increasingly weaker as one moves away from
the   center  of   the  band,   mostly  because  the   single-particle
localization length decreases.  Hence, for smaller values of $E_F$, it
would be  difficult to identify   a suppression of the  delocalization
effect relative to the two-particle effect.

In Fig.~2 we exhibit our results for the  two pair energies $E=-2$ and
$E=0$  corresponding  to    excitation   energies  $\epsilon=0$    and
$\epsilon=2$, respectively.   We   plot  $\xi_q(E)/\xi_1(E/2)$  as   a
function  of  $u\xi_1(E/2)/t$ analogous  to  the scaling  plot for the
two-particle   problem.\footnote{Note that  the        single-particle
  localization energy should be evaluated at half the pair energy.  We
  generated    the  single-particle   localization lengths   for these
  energies  numerically  by   the  recursive  Green   function  method
  \cite{Huckestein} using  systems with $10^8$  sites.  } While  we do
not find the same scaling any  more in the many-particle problem, this
still allows us to exhibit data for all  values of the interaction and
the  disorder strength in  a single plot.  Clearly, the delocalization
effect     disappears   almost  completely     for  zero    excitation
energy.\footnote{Note    that   we     find      $\xi_2<\xi_1/2$   for
  $\epsilon\!=\!0$.   This is  a  consequence of  neglecting the first
  factor in Eq.~(\ref{green}) \cite{Oppen}.   However, there is indeed
  no  delocalization  effect in this case,  because  the range of both
  factors in  (\ref{green}) is independent  of  $u$ and $W$.}  On  the
other  hand, the delocalization of  the quasiparticle pair is close to
that  of  the  two-particle  problem for    $\epsilon=2$  as seen   by
comparison  with  the two-particle result  at   $E=0$ also  plotted in
Fig.~2.
\begin{figure}
\centerline{\psfig{figure=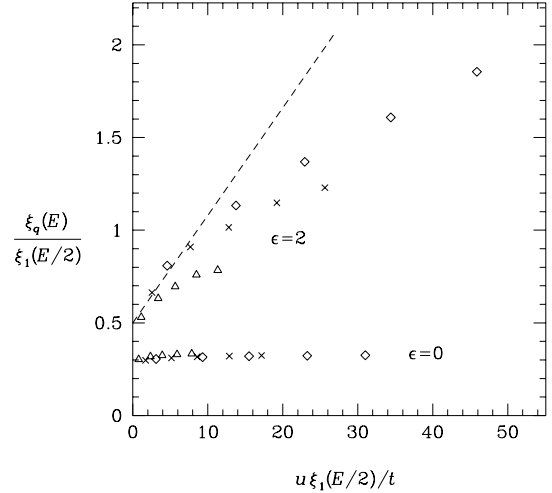,width=75mm}}
\caption{Plot of $\xi_q(E)/\xi_1(E/2)$ vs $u\xi_1(E/2)/t$ for two 
  different excitation energies.  Five  values of $u$ are included for
  each   of the three  values  of  disorder  $W=3$ (triangles),  $W=2$
  (crosses),  and $W=1.5$ (diamonds).  The  two-fermion effect is also
  shown for  comparison (dashed line).   Evidently, the delocalization
  effect  has disappeared  at the  Fermi energy  and is recovered with
  increasing excitation energy.  One also observes deviations from the
  scaling behavior of the two-fermion case.}
\end{figure}

In view of the absence of the delocalization effect at zero-excitation
energy, it  is important to identify  the relevant scale  on which the
effect is recovered with increasing excitation energy.  To investigate
this question, we have considered  the enhancement of $\xi_q$ relative
to the enhancement of $\xi_2$ for the two-particle problem as measured
by
\begin{equation}
  R_{u,W}(\epsilon)
  ={\xi_q(u)-\xi_q(u=0.1)\over\xi_2(u)-\xi_2(u=0.1)}\;.
\label{effect}
\end{equation}
We have computed this quantity as a function  of the excitation energy
$\epsilon$ for different  disorder and interaction strengths.  Between
$\epsilon\!=\!0$  and  $\epsilon\!=\!2$,    the quantity should   vary
roughly   between  zero  and   one.  Figs.~3(a)   and   (b) show  that
$R_{u,W}(\epsilon)$ at  fixed   $u$ is   independent of the   disorder
strength    $W$.\footnote{The     accuracy    of    the     data   for
  $R_{u,W}(\epsilon)$   is  limited    because both  numerator     and
  denominator in   (\ref{effect})  are small  numbers.}   Moreover,  a
comparison of Figs.~3(a) and (b) indicates that $R_{u,W}(\epsilon)$ is
also  independent of the  interaction strength $u$.  This implies that
the bandwidth $t$ is the  relevant scale  on which the  delocalization
effect is recovered with  increasing excitation energy. This result is
in  qualitative  agreement  with   the prediction  from  Thouless-type
arguments by Imry \cite{Imry}.
\begin{figure}
\centerline{\psfig{figure=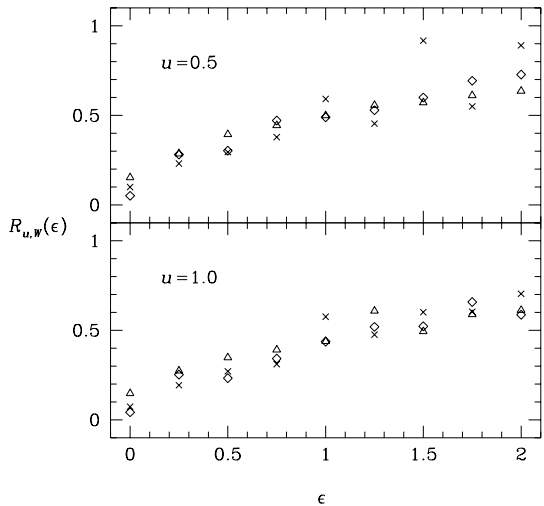,width=75mm}}
\caption{Plot of the delocalization effect $R_{u,W}(\epsilon)$ as 
  defined in   Eq.~(\protect\ref{effect}) as a  function of excitation
  energy  for  two values  of $u$ and  three  values of disorder $W=3$
  (triangles), $W=2$ (crosses),   and $W=1.5$  (diamonds).   Note that
  $R_{u,W}(\epsilon)$ is insensitive to  both disorder and interaction
  strength.}
\end{figure}

Our   results  were obtained using   a   simple approximation  to  the
two-particle Green function. Even within improved approximations we do
not expect a delocalization effect analogous to that for two particles
to reappear at low excitation energies because  the suppression of the
two-particle effect  at finite  density  is entirely due to  the Pauli
principle.  This   expectation  is also     in accord with    standard
phase-space  arguments for the  degenerate Fermi system.  On the other
hand, our approximation does  not allow us  to exclude the possibility
that the localization  properties of    the Anderson insulator     are
strongly affected by true correlation effects.

We  conclude by discussing the physical  implications  of our results.
For  thermodynamic and  low-frequency  transport properties  in linear
response,  the relevant excitation  energies are of  the  order of the
temperature.    In  these  cases,    the  maximal  delocalization   of
quasiparticle pairs is {\it not\/} expected to be observable since the
bandwidth will generally be much larger than the temperatures at which
quantum-interference    effects  such as   Anderson  localization  are
observable.  In  this  sense,  the  quasiparticle   delocalization  at
finite density is  significantly weaker than the  two-particle effect.
However,   it may   still  be  possible    to observe the    incipient
delocalization  at excitation energies  which  are low compared to the
bandwidth.  In particular,  this may be  possible in higher dimensions
\cite{Imry} where  the delocalization  effect  is expected to  be more
pronounced and   where  the  single-particle  localization  length can
become exceedingly large   (e.g., close to   the Anderson transition).
Finally, we note that there may also be a possibility of detecting the
delocalization of quasiparticle pairs in high-frequency experiments.

We  enjoyed  helpful discussions with  G.\ Barkema,  J.\ M\"uller, and
H.A.\ Weidenm\"uller.

\end{document}